\documentclass[aps,prl,twocolumn,groupedaddress,showpacs]{revtex4-1}
 \usepackage{graphics}

\begin{document}


\title{A nonmetric theory of gravitation that is nonsingular at the Schwarzschild radius}


\author{James W. Douglass}
\email[]{james.douglass@cfl.rr.com}

\noaffiliation

\date{\today}

\begin{abstract}
A gravitational theory is formulated by considering the physical processes underlying relativistic dilation of time and contraction of space.  It is shown that the point mass solution of general relativity's field equation - the Schwarzschild metric - is a weak-field approximation to the more general Lagrangian resulting here. Unlike general relativity the resulting theory does not exhibit a singularity at the Schwarzschild radius and furthermore shows that photons escape from that radius with a red shift of $e^{\frac{1}{2}}-1$.  Consequently black holes are not black.  Experimental confirmations of general relativity that have been conducted in the weak field domain may also be considered as confirmations of the theory presented here since the two theories provide nearly identical predictions there.  Testable differentiation of these two theories will require observations in the near vicinity of the Scnwarzschild radius of a black hole.   Issues related to experimental confirmation are discussed.  
\end{abstract}


\maketitle

\section{1. INTRODUCTION}  

This paper presents a relativistic theory of gravitation that is developed by considering the concrete physical mechanisms responsible for time dilation and spatial contraction.  The theory is developed for the central gravitational field of a non spinning mass.   The mathematical framework is formulated from thought experiments in which the inverse square acceleration of Newtonian gravitation and the principles of relativity are combined.   The Euler Lagrange equations based on the resulting Lagrangian for extremal proper time are shown to be equivalent to the geodesic equations of general relativity with the Schwarzschild metric.   However, unlike general relativity this theory is nonsingular at the Schwarzschild radius and therefore uniquely provides a description of particle dynamics there.   The Schwarzschild metric of general relativity is shown to be an first order approximation to the Lagrangian developed here.

Relativistic principles, terms, and concepts used herein are given below with comments and references where appropriate.  Since numerous references exist in both classical and modern publications for these only a sampling is provided where indicated.  They are identified here as essential logical building blocks of the sections that follow.

\begin{itemize}

\item The principle of equivalence - the original version of this principle given by Einstein  \cite{Einstein} equated gravitational and inertial masses.   Since that time a number of logically related statements of equivalence have appeared \cite{Adler} \cite{Carroll} \cite{Misner} \cite{Weinberg}.   For purposes of this paper we will take the meaning of this principle to include the equivalence of gravitational and inertial masses, independent of their nature and we also utilize the ``strong'' version of equivalence \cite{Hobson} stating that a laboratory that is freely falling in a gravitational field is equivalent to one in an inertial, or Lorentz frame.   A mental tool used to conduct thought experiments will be an elevator, an Einstein elevator acted on only by gravitational forces and in free fall.   This principle therefore states that results of experiments conducted entirely within the elevator be identical to those conducted in an unaccelerated frame of reference - ignoring tidal effects caused by spatial gradients of the gravitational field.  The Lorentz transformations \cite{Hobson} will be applicable within this elevator.

\item The mass-energy equivalence of special relativity - will be used to associate a mass of an object to its total energy.   Objects may be photons.  

\item The weak field approximation - A parameter that appears in the following theory as well as in general relativity is chosen to quantitatively establish the gravitational field ``strength'.  At this point it is easiest to extract this parameter from the Schwarzschild metric, which in  polar coordinates is given by \cite{Weinberg2} \cite{Walecka}:

 \begin{equation}
  \label{Schwarzschild_metric} 
 {ds}^2=c^2{dt}^2{(1-\frac{r_s}{r}}) -\frac{{dr}^2}{(1-\frac{r_s}{r})}-r^2{d\theta}^2-r^2{\sin\theta}^2{d\phi}^2
 \end{equation}

 Here, ${ds}$ is the space-time interval,  $r$ is the radial coordinate of a test mass in a spherical coordinate system centered on a mass $M$,  $dr$ its differential, $d{\theta}$ and $d{\phi}$ are the angular coordinate differentials, $r_s$ is the Schwarzschild radius calculated for a mass M as $r_s=\frac{2GM}{c^2}$, $G$ is the gravitational constant, and $c$ is the speed of light.    Within general relativity the variation of the integral of $ds^2$ over a path length is minimized, or in other words particles follow geodesic paths in a warped space-time.   The effects of space-time warping, and therefore observable differences in predictions in general relativity and Newtonian gravitational theories are quantified by the ratio $\frac{r_s}{r}$ in the Scwarzschild metric above.  The weak field limit is defined here as the regime where $\frac{r_s}{r}<<1$.  For use later in this paper, the gravitational radius will be defined as one half the Scwarzschild radius, and the weak field definition applies as well to this radius.   Original and subsequent experimental  tests of general relativity using the sun as the central gravitational body (light deflection, orbital perturbations, red shifting) are necessarily in the weak field regime as defined above.   Consider gravitational deflection of light passing the sun's limb; since the physical radius of the sun is on the order of 700,000 km, and the calculated Schwarzschild radius approximately 3 km, the ratio of $\frac{r_s}{r}$ is on the order of $3/700000$, or about 4 parts per million.  This ratio's value is even smaller for planetary distances and confirmations based on those (e.g., perihelion advance of Mercury).  
 
 \item Newtonian gravitation -  this law describes gravitational effects as a radial force (no azimuthal component) that varies inversely with the square of the distance from a mass $M$ to a test mass $m$ which is assumed small relative to $M$.  While the radial-only nature of this law is retained here, it will be viewed as an inverse-square \textit {acceleration law}, applying to any mass regardless of its type.
 
 \item Paths of extremal proper time -  in this paper, this principle refers to requiring particles in a gravitational field to follow paths resulting in variationally stationary, or extremal proper time. This principle is analogous to seeking geodesic paths in space time \cite{Hobson2}.   However we wish to utilize this principle a-priori and independently of any other theory.  
 
Non relativistic physics provides examples  of such a principle including Fermat's least time description of photon propagation as well as the least action formulation by Maupertuis   \cite{Goldstein}.   
 
That paths are those of extremal time may also be seen from Maupertuis's least action principle in another, semi-classical way.   Seeking paths that minimize the action given by:
 
 \[
 \Delta\int_{path} p \ dq=0
\]       

where $\Delta$ refers to the form of variation of the integral where the endpoints may also vary, $p$ the particle's momentum, and $q$ a generalized coordinate.   If we associate a quantum mechanical wavelength $\lambda$  with this particle according to $\lambda=\frac{h}{p}$, where $h$ is Plank's constant,  substitution into the equation above results in (ignoring the multiplicative constant $h$ since it does not affect the final result of the variation):

\[
\Delta\int_{path}\frac{dq}{\lambda}=0
\]

The reciprocal of $\lambda$ may be viewed as the number of oscillations, or cycles of time the particle experiences, it's proper time, per unit of $q$.  The equation above then yields paths of stationary proper time. 

All of these results may be made relativistic by the principle of equivalence.  By this principle an observer in an accelerating frame will identify the same particle path as a (proper) elapsed time extremal as would an observer in a non-accelerating frame.  

\end{itemize}

The development of the theory presented here emphasizes visualization of physical processes through the use of thought experiments.  Transformations are derived relating time and spatial coordinates in a non accelerating reference frame $K$ to those in a freely falling  frame of reference $K^\prime$.   A third frame is then introduced, $K^{\prime\prime}$ which is in uniform motion with respect to $K^\prime$.  The Lorentz transformation is utilized to transform time and spatial coordinates from  $K^\prime$ to $K^{\prime\prime}$, and then ultimately to $K$.  These result are then used to produce an expression for an increment of time in the accelerating and moving frame $dt^{\prime\prime}$ which becomes the Lagrangian for arriving at paths of stationary proper time generated by:

  \[\delta\int_{path}dt^{\prime\prime} = 0.  \]
  
It is then shown that this Lagrangian reduces to Eq.(\ref{Schwarzschild_metric}), the Schwarzschild metric in the weak field limit and therefore provides an equivalent description of gravitation as general relativity in this regime.  It is further shown that this Lagrangian provides non-singular solutions at the Scnwarzschild radius as well as over all radii (except at the coordinate origin).  

\section {2. THE TRANSFORMATIONS}

\subsection {A. The dilation of time due to gravitational acceleration}
In special relativity one may generate time and spatial transformations between a reference and uniformly moving frame by considering a Michaelson interferometer with one arm oriented parallel to the velocity vector and the other arm oriented 90 degrees from that.   In such an approach the paths light follows along these two arms are calculated from the perspective of the stationary frame of reference, and then by utilizing the principles of relativity and the postulate of the constancy of the speed of light one obtains both the temporal and spatial transformations.   

At first it would appear this approach would be suitable for developing transformations between a non accelerating and an accelerating frame of reference.   However an accurate description of the path light in a gravitational field, particularly a path involving both radial and transverse motion is indeed a major goal of this paper.   A different approach will therefore be taken based on photons in free fall as presented below.  

We will make use of an Einstein elevator' which is in free fall in a gravitational field, i.e. the only force acting on this elevator and its contents is gravitation.  An observer inside this elevator uses an atomic clock to measure time in his frame of reference which we will designate as $K^\prime$.  While an atomic clock is clearly only one possible way to measure time, it is based on fundamental processes and so will be considered a manifestation of time itself.

As observed from $K$, $K^\prime$ is accelerating in the radial direction only toward the center of mass of $M$ as are photons within $K^\prime$.  A test mass $m$ will undergo a change in  energy $dU$ due to an incremental change in radius $dr$ according to:

\begin{equation}
\label{potential}
dU=\frac{GMm}{r^2}dr
\end{equation}

where $G$ is the gravitational constant, $r$ is the radial coordinate of the test mass in a spherical coordinate system centered on the center of mass of $M$.  It is assumed that $m$$\ll$$M$.  

We may associate a mass with a photon emitted by the atomic clock according to Einstein's mass-energy equivalence $E=mc^2$ where $E$ is the photon energy, and $c$ is the speed of light.  Writing the photon energy as the sum of its energy at $r=\infty$ and its potential energy at a finite $r$, we have for its mass:

\begin{equation}
m=\frac{h\nu_0+U}{c^2}
\label{mass}
\end{equation}

where $\nu_0$ is the photon frequency at $r=\infty$, and $h$ is Plank's constant. Substitution of this result into Eq. (\ref{potential}) gives:
\[
dU=GM\left (\frac{h\nu_0+U}{c^2}\right )\frac{dr}{r^2}
\]
\[
\frac{dU}{h\nu_0 + U}=\frac{GM}{c^2}\frac{dr}{r^2}
\]

Defining the gravitational radius $r_g=\frac{GM}{c^2}$ and using the condition that $U=0$ at $r=\infty$, we have for $U$ :

\[
U=h{\nu_0}\left (e^{-\frac{r_g}{r}}-1\right )
\]
Since the energy of the photon is the sum of it's energy at $r=\infty$ and $U$, and since it is also given by $h\nu^\prime$, we have:

\begin{eqnarray*}
 E &=& \ h\nu_0 + U  \qquad      \mbox{ Substituting for $U$ from above:}\\
 E &=& h{\nu_0}\ e^{-\frac{r_g}{r}} \qquad \  \mbox{and since also:}\\
E &=& \ h \nu^{\prime}  \qquad \qquad \  \mbox{we have:}\\
 h{\nu^{\prime}} &=& h{\nu_0}\ e^{-\frac{r_g}{r}}
 \end{eqnarray*}
 
Therefore:

\begin{equation}
\nu^{\prime}={\nu_0}\ e^{-\frac{r_g}{r}}
\label{red_shift}
\end{equation}

This expression indicates the emission frequency $\nu^{\prime}$ at a radial coordinate of $r$ is lower than when at $r=\infty$, and has a value of zero at $r=0$.  This lowering of frequency is otherwise known as gravitational red shifting. 

An observer in $K$ may determine the frequency of these photons as the number of wave crests $N$ passing a fixed location per unit of time.   From Eq. (\ref{red_shift}) above above he notices that an observer in $K^\prime$ will observe a reduced number of crests in that same unit of his time owing to its reduced frequency.   However the principle of equivalence requires observers in $K^\prime$ measure the same frequency as in $K$.  Therefore the rate of time in $K^\prime$ must have changed according to :

\[
\frac{dN}{dt} = \frac{e^{-\frac{r_g}{r}}dN}{dt^\prime}=\frac{e^{-\frac{r_g}{r}}dN}{e^{-\frac{r_g}{r}}dt}
\] 

\begin{equation}
\label{ttransform}
dt^\prime = e^{-\frac{r_g}{r}}dt
\end{equation}

This is the transformation for a time differential  between  $K^\prime$ and  $K$. 

\subsubsection{1. The spatial dependence of gravitational time dilation}

 One of the teachings of special relativity is that there is a time dilation effect due to uniform motion on synchronization of spatially separated clocks.   This effect is sometimes known as the relativity of simultaneity.  We now evaluate whether this mechanism results in a significant time dilation effect in a free-falling frame. 
 
\begin{figure}[t]
\graphicspath{ {Figures/} }
\scalebox{0.5}
{\includegraphics{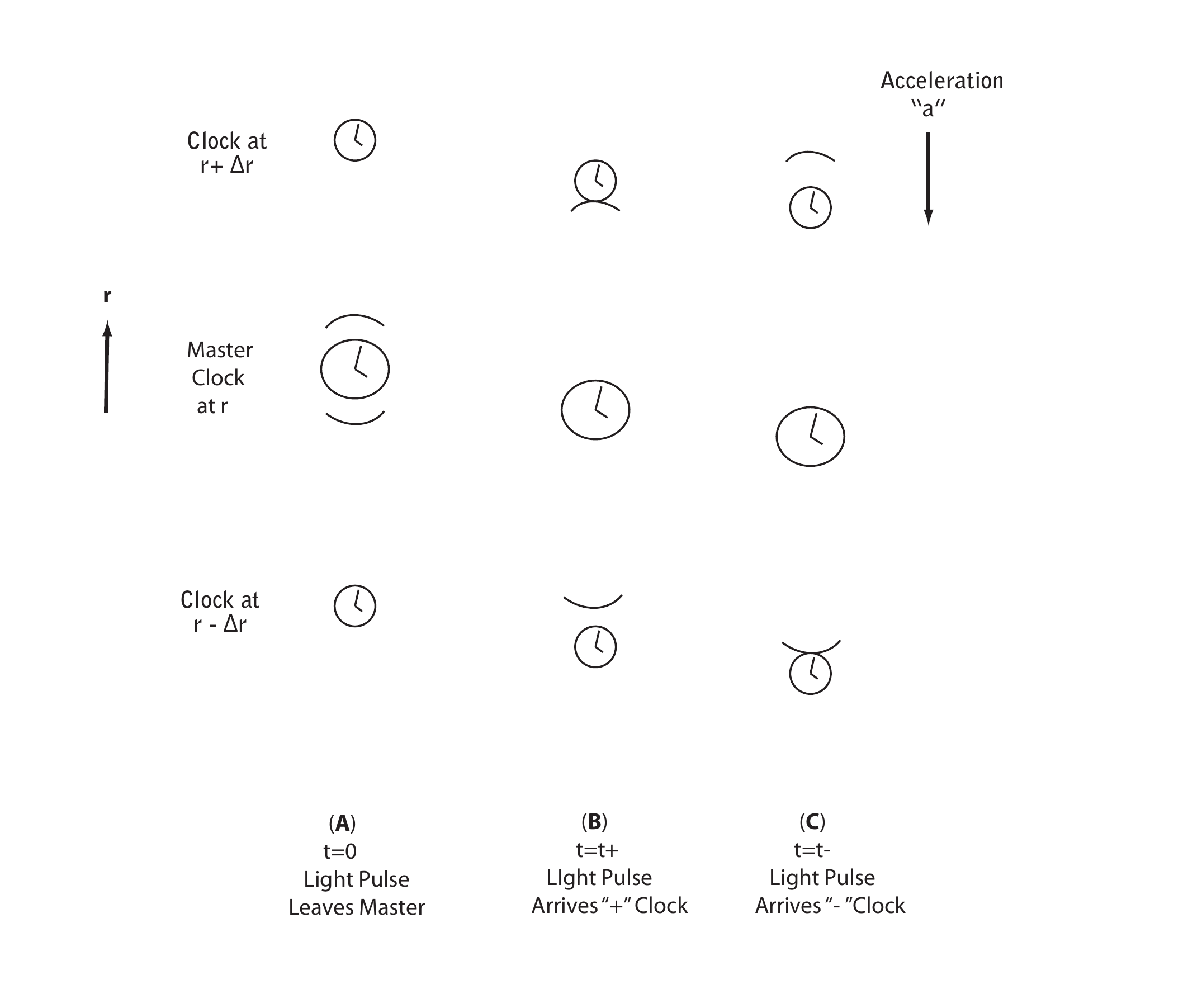}}
\caption{Clock Synchronization in $K^\prime$ as Seen From $K$}
\label{SpatialTime}
\end{figure}

   Assume that spatially separated clocks in $K^\prime$ are synchronized by the transmission of light pulses from a single clock designated as the master clock.   Upon reception of these timing pulses each of the other clocks within $K^\prime$ adjusts their clocks to the previously agreed to time of transmission of these pulses after accounting for their respective propagation delays.   As seen from $K$, however the reception time of a reference pulse is not only determined by the propagation time, but also by an additional effect due to acceleration.  This is illustrated in Figure \ref{SpatialTime}.

     The (A) portion of this Figure illustrates how a timing pulse is transmitted at time $t=0$ by the master clock located at a radial coordinate $r$, to a remote clock located at a radial coordinate of $r+\Delta r$.  The (B) portion of this figure shows that at the time the pulse is received, $t_+$  the remote clock's radial position has decreased by $\frac{a}{2}({t_+})^2$ due to acceleration, $a$.  This causes the timing pulse to be received earlier than by considering the propagation time alone.   Since in time $ t_+$ light has propagated a distance of $ct_+$, we have:
     
     \[ ct_+=\Delta r - \frac{a}{2}(t_+)^2\]
    
     \[  t_+^2+\frac{2c}{a} t_+ - \frac{2 \Delta r}{a}=0\]
     
     The solution for $t_+$  is given by:
     
     \begin{equation}
     t_+=\frac{c}{a}\Bigg [ -1 \pm \sqrt{1+\frac{2a\Delta r}{c^2}} \Bigg ]      
     \label{time received}
     \end{equation}

   We may define the factor describing the reduction in arrival time compared with the propagation time alone as
\begin{equation}
\Delta t_+ =\frac{\Delta r}{c} - t_+
\label{delta t}
\end{equation}

 Expanding the radical term in the Eq.(\ref{time received}) we have:
 \[
\Delta t_+ = \frac{\Delta r}{c} - \frac{c}{a}\Bigg [ -1 \pm \bigg (1+\frac{a \Delta r}{c^2}-\frac{1}{8}{\bigg({\frac{2a\Delta r}{c^2}}\bigg )}^2+...\bigg )\Bigg ]
\]
For a positive $\Delta r$ as shown in the figure above we may choose the $+$ sign in the expanded radical, and retaining to second order:
\[
\Delta t_+ \simeq  \frac{\Delta r}{c} - \frac{\Delta r}{c}+\frac{a{\Delta r}^2}{2c^3} = \frac{a{\Delta r}^2}{2c^3}
\]
  
Substituting $a=\frac{GM}{r^2}$, and $r_g=\frac{GM}{c^2}$  we have:

\[
   \Delta t_+  \simeq \frac{r_g {\Delta r}^2}{2cr^2}
\]

The (C) portion of the above figure shows a similar effect for the clock located at $r-\Delta r$.   The same analysis as the above shows this clock is synchronized to a time that is greater than that in the absence of acceleration by the factor $\Delta t_-$ that is given by:

\[
   \Delta t_-  \simeq \frac{r_g {\Delta r}^2}{2cr^2}
\]

And so we see that the spatial (radial) dependence in time due to synchronization effects  is second order in $\Delta r$, and so can be ignored compared with the first order time dilation effect  represented by Eq. (\ref{ttransform}).  Spatial dependence of synchronization  in the angular (orthogonal) coordinates are zero since acceleration is zero there.

\subsection{B. The contraction of space due to gravitational acceleration}  
As before we will utilize a spherical coordinate system centered on the center of mass of $M$.  We once again consider a closed elevator located at a radial coordinate $r$ whose acceleration (and that of its contents) toward the the origin  is given by $a=\frac{GM}{r^2}$.  The effect of this acceleration on a measurement of distance is presented below.

\subsubsection{\textit{1. Radial transformation}}
\begin{figure}[t]
\graphicspath{ {Figures/} }
\scalebox{0.4}
{\includegraphics{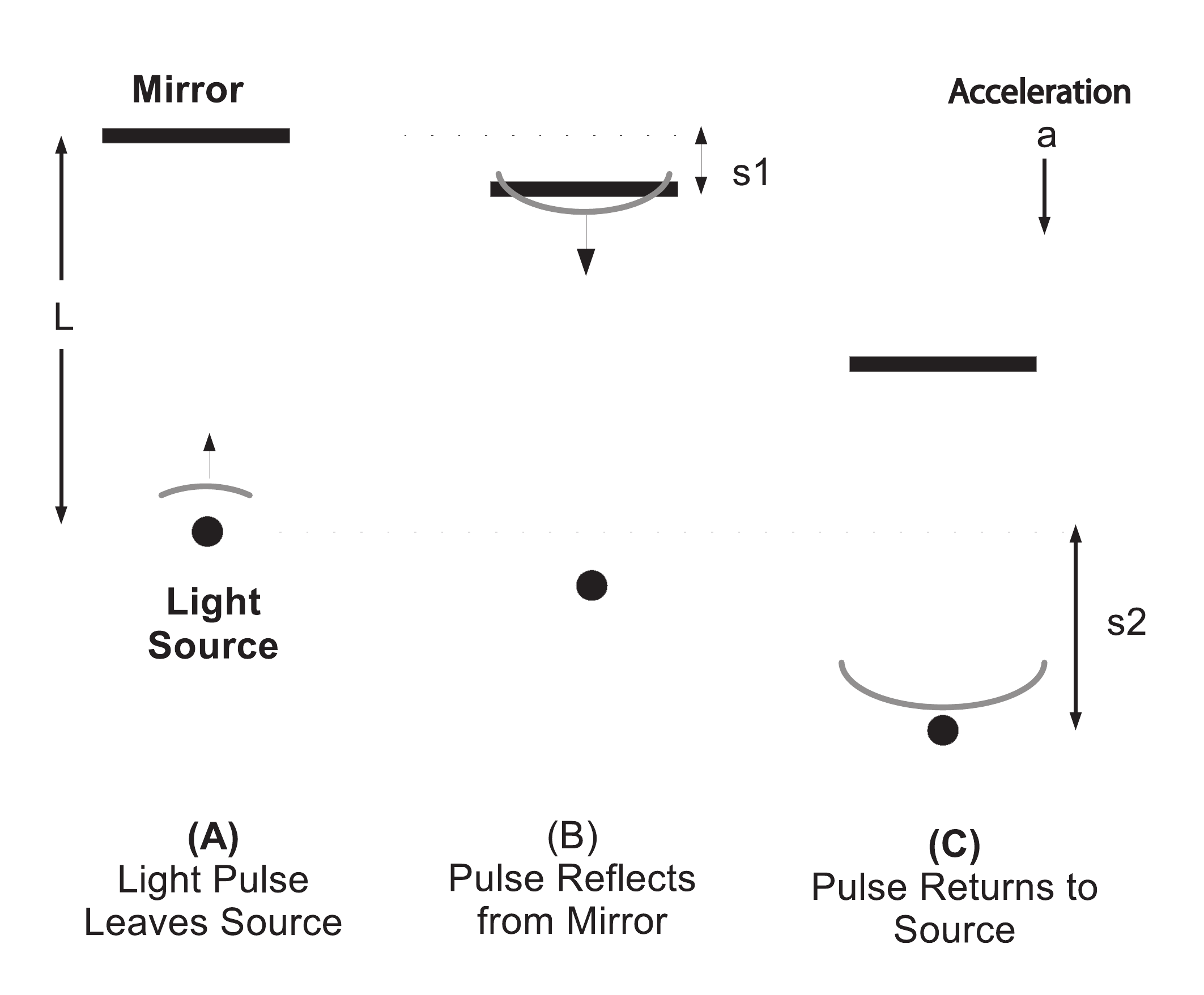}}
\caption{A Measurement of Length in $K^\prime$ as Seen from $K$ }
\label{Transforms}
\end{figure}

Consider a thought experiment in which an observer located within an elevator measures the speed of light with a light source and reflecting mirror oriented  in the radial direction as shown in Figure  \ref{Transforms}.  He measures the distance between the source and mirror $L^\prime$  using a metering rod provided by an observer in $K$ who measured its length prior to acceleration as $L$.  

The measurement is performed by directing a light pulse from a source toward a mirror as shown in Figure \ref{Transforms}A.  As observed from $K$ the mirror translates  a distance  s1 during the travel time of the light pulse that is given approximately by $\frac{L}{c}$.    The pulse is reflected and returns to the source's location by which time it has translated a total distance of s2 (Figure \ref{Transforms} B and C).   

We have for the total upward and downward distances traveled by the light pulse given respectively by:
\[
s_{up}=L-s1\qquad\text{and}\qquad s_{down}=s_{up}+s2
\]

\[
s_{up}\simeq L-\frac{a}{2} \left(\frac{L}{c}\right)^2, 
\qquad\text{and}\qquad
 s_{down}\simeq s_{up}+\frac{a}{2} \left(\frac{2L}{c}\right)^2
\]
\[
s\equiv s_{up}+s_{down}=2\left[L-\frac{a}{2} \bigg(\frac{L}{c}\bigg)^2\right]+\frac{a}{2} \left(\frac{2L}{c}\right)^2
\]
\begin{equation}
s=2L+a\bigg(\frac{L}{c}\bigg)^2
\end{equation}

 If the time required for light to travel $s$ is $\Delta t$ we have in frames $K$ and $K^\prime$:
\[
c=\frac{s}{\Delta t} =\frac{\bigg[2(L)+a\bigg(\frac{L}{c}\bigg)^2\bigg]}{\Delta t}=\frac{2L^\prime}{{\Delta t}^\prime}
\]   
     
Substituting the relationship between time increments from Eq (\ref{ttransform}), $a=\frac{GM}{r^2}$, and $r_g=\frac{GM}{c^2}$ and then solving the above equation for $L^{\prime}$, the length of a metering rod in $K^\prime$ is:
     
\[
L^\prime= e^{-\frac{r_g}{r}}\left(L+\frac{r_gL^2}{2r^2}\right)
\]

  We see again that the effect of acceleration on the measurement of a standard rigid rod of length $L^\prime$ is second order.   Therefore this second order factor can be ignored.  This is especially true since a standard of length should be ``small'' so as to afford precision in the measurement of lengths.  The expression above therefore reduces to:

\[
L^\prime= e^{-\frac{r_g}{r}}L
\]

Since radial distance, as well as an incremement of radial distance  is the number of standard metering rods given by $L^\prime$, we have  that:
\begin{equation}
\label{rtransform}
dr^\prime=\frac{dr}{e^{-\frac{r_g}{r}}}
\end{equation}

 This expression therefore  is the desired transformation for radial differentials between frames $K$ and$K^\prime$.  
 
 \subsubsection{\textit{2. Azimuthal transformations}}
 The same methodology applied to dimensions orthogonal to $r$, i.e. $ \phi$ and $ \theta$, where acceleration is assumed to be zero, and the principle of relativity yields:
 
\begin{equation}
\label{angular_transforms}
d\phi^\prime = d\phi,
\qquad\text{and}\qquad
d\theta^\prime=d\theta
\end{equation}

\subsection {C. The effect of uniform motion}
     The previous sections have addressed time and spatial transformations due to gravitational acceleration only.  The effects of uniform motion will now be added.   We do so by introducing the frame of reference $K^{\prime\prime}$ which is also in free fall and in uniform motion with respect to $K^\prime$ with velocity $v^\prime$.  Since frame $K^\prime$ is stationary with respect to $K$, $v^\prime$ = $v$.  
     
     While the effects of uniform motion (special relativity) could be introduced from first principles, it is more efficient to utilize the Lorentz transformations which must be valid in $K^\prime$ according to the equivalence principle .   Taking the special relativistic time transformation between $K^{\prime\prime}$ and $K^\prime$ as:
     
\begin {equation}
\label {dt_double_prime}
dt^{\prime\prime}=  { \left[1-\frac{{v^\prime}^2}{c^2}\right]}^\frac{1}{2}dt^\prime
\end {equation} 
  In polar coordinates this may be written as:

\[
dt^{\prime\prime}=  { \left[1-\frac{1}{c^2}{\left[{\frac{dr^\prime}{dt^\prime}}^2+r^2\left({\frac{{d\theta}^\prime}{dt^\prime}}^2   +   \sin^2\theta\,{\frac{d\phi^\prime}{dt^\prime}^2}\right)    \right]}\right]}^\frac{1}{2}dt^\prime
\] 

\[
dt^{\prime\prime}=  { \left[{dt^\prime}^2-\frac{1}{c^2}{\left({{dr^\prime}}^2+r^2\left({{{d\theta}^\prime}}^2   +   \sin^2\theta\,{{d\phi^\prime}^2}\right)     \right)}\right]}^\frac{1}{2}
\]
For brevity we substitute for the angular dimension (orthogonal to  radius) ${d\Omega^\prime}^2=r^2\left({d\theta^\prime}^2 + \sin^2  \theta \,{d\phi^\prime}^2\right)$.  The equation above becomes:

\begin{equation}
\label{double_prime_time}
dt^{\prime\prime}=  { \left[{dt^\prime}^2-\frac{1}{c^2}{\left({{dr^\prime}}^2+{d\Omega^\prime}^2 \right)}\right]}^\frac{1}{2}
\end{equation}

The transformation to $K$ is made by substituting for $dt^\prime$, $dr^\prime$, $d\theta^\prime$, and $d\phi^\prime$ from Eqs (\ref{ttransform}),(\ref{rtransform}), and (\ref{angular_transforms}) to obtain:
 
\[
dt^{\prime\prime}=  { \left[e^{-\frac{2r_g}{r}}\,{dt}^2
-\frac{1}{c^2}
\left(\frac{dr^2}{e^{-\frac{2r_g}{r}}}
+{d\Omega}^2
\right)
\right]
}^\frac{1}{2}
\]
Substituting $r_g = \frac{1}{2}r_s$ where $r_s$ is the Schwarzschild radius, we finally arrive at an expression for the proper time increment in a frame which is both accelerating and in uniform motion:

\begin{equation}
\label{dt_double_prime}
dt^{\prime\prime}=  { \Bigg[e^{-\frac{r_s}{r}}\,{dt}^2
-\frac{1}{c^2}
\left(\frac{dr^2}{e^{-\frac{r_s}{r}}}
+{d\Omega}^2
\right)
\Biggr]
}^\frac{1}{2}
\end{equation}

\section {3. THE RESULTING EQUATIONS OF MOTION}

From the preceding sections, paths resulting in stationary proper time are given by:

\begin{equation}
\label{variation}
\delta\int_{path}{dt^{\prime\prime}} =\delta\int_{path}{\left(1-\frac{{v^\prime}^2}{c^2}\right)}^\frac{1}{2}dt^\prime=0
\end{equation}
Differential equations of motion may therefore be obtained by using  Eq.(\ref{dt_double_prime}) as a Lagrangian with substitution into the Euler Lagrange equation, or:

\begin{equation}
\label{Lagrangian}
L={ \Bigg[e^{-\frac{r_s}{r}}\,{dt}^2
-\frac{1}{c^2}
\left(\frac{dr^2}{e^{-\frac{r_s}{r}}}
+{d\Omega}^2
\right)
\Biggr]
}^\frac{1}{2}
\end{equation}
with:

\[
\frac{d}{dt}\frac{\partial {L}}{\partial{\dot{q}}}-\frac{\partial {L}}{\partial{q}}=0
\]
Where $q$ is a generalized coordinate, and dot refers to time differentiation.

Once again it is important to recall the unit analysis results presented in the Radial transformation section regarding the true units of the denominator in this expression.

 Examples of the resulting differential equations are not provided here due to space limitations and mathematical complexity, however limiting cases are compared with general relativity in the following section.

\section{4. COMPARISON WITH GENERAL RELATIVITY}

As shown previously the Lagrangian of Eq. (\ref{Lagrangian}) provides equations of motion resulting from this theory.   We will show that the square of this Lagrangian provides the same equations of motion.   We will then show that this squared form of the Lagrangian reduces to the Schwarzschild radius in the weak field limit.  

  Recall that the frames of reference $K^\prime$ and $K^{\prime\prime}$ are both in free fall with respect to  $K$, and that $K^{\prime\prime}$ is in uniform motion with respect to $K^\prime$ with velocity $v^\prime$.  Consider a particle which is at rest in $K^{\prime\prime}$.  As seen from $K^\prime$ this particle will be in uniform motion with speed $v^\prime$.  An observer in $K^\prime$ who wishes to describe the motion of this particle would write a Lagrangian as the particle's constant kinetic energy only resulting in the variational equation:

\[
\delta\int\frac{1}{2}m{v^{\prime}}^2 dt^\prime=0
\]

where $m$ is the particle's mass.   Solutions to this equation are not changed by a multiplicative constant.     Without loss of generality we may therefore replace the constant $\frac{m}{2}$ with $\frac{1}{c^2}$ to obtain:

\begin{equation}
\label{KE Only}
\delta\int\frac{{{v^\prime}}^2}{c^2} dt^\prime=0
\end{equation}

And since the elapsed proper time for this particle over its path is given by:

\[
\int_{t_i^\prime}^{{t_f}^\prime}dt^\prime={t_f}^\prime-{t_i^\prime}
\]
which is a function of $t_i^\prime$ and $t_f^\prime$ only.  We choose the variational formulation such that the end points are fixed and hence we may write:

\begin{equation}
\label{proper prime}
\delta\int{dt^\prime}=0
\end{equation}

Taking the difference of Eqs. (\ref{proper prime}) and (\ref{KE Only}) we obtain:

\begin{equation}
\delta\int{dt^\prime}-\delta\int\frac{{{v^\prime}}^2}{c^2} dt^\prime=0 \mbox{    ,or collecting terms}   
\end{equation}

 \begin{equation}
 \label{Variation prime}
 \delta\int\left(1-\frac{{v^\prime}^2}{c^2}\right)dt^\prime=0
\end{equation}

Substituting:
\begin{equation}
\label{Lagrangian prime}
L^\prime={\left(1-\frac{{v^\prime}^2}{c^2}\right)}^\frac{1}{2}
\end{equation}

Eq. (\ref{Variation prime}) becomes:
\begin{equation}
\label{Lagrangian prime squared}
\delta\int{{L^\prime}}^2dt^\prime=0
\end{equation}

Note that the integrand in Eq. (\ref {Variation prime}) is exactly the square of that in Eq. (\ref{variation}) which is the variation of elapsed proper time.   We will now show that these two equations are equivalent for the generation of equations of motion.

Differential equations of motion may be obtained from  Eq. (\ref{Lagrangian prime squared}) from the Euler Lagrange equation to obtain, where the dot notation refers to differentiation with respect to $t^\prime$:
\[
\frac{d}{dt^\prime}\frac{\partial{({L^\prime}^2})}{\partial{\dot{q^\prime}}}-\frac{\partial{({L^\prime}^2})}{\partial{{q^\prime}}}=0
\]

\[
\frac{d}{dt^\prime}(2L^\prime)\frac{\partial{{L^\prime}}}{\partial{\dot{q^\prime}}}-(2L^\prime)\frac{\partial{{L^\prime}}}{\partial{{q^\prime}}}=0
\]

     Since the particle is moving with constant speed in $K^\prime$, $L^\prime$ is constant and may be factored to leave:
\[
\frac{d}{dt^\prime}\frac{\partial{{L^\prime}}}{\partial{\dot{q^\prime}}}-\frac{\partial{{L^\prime}}}{\partial{{q^\prime}}}=0
\]

This is precisely the Euler Lagrange equation that would follow from Eq. (\ref{variation})  demonstrating the equivalence of Eqs. (\ref{variation}) and (\ref{Variation prime}).            
    
We now transform the previously defined Lagrangian:

\begin{equation}
L^\prime={\left(1-\frac{{v^\prime}^2}{c^2}\right)}^\frac{1}{2}
\end{equation}

to Frame $K$ using the transformations from  Eqs (\ref{ttransform}),(\ref{rtransform}), and (\ref{angular_transforms}) and substitute into Eq. (\ref{Lagrangian prime squared})  to obtain:

\begin{equation}
\label{full}
\delta\int_{path}
{\left[
 e^{-\frac{r_s}{r}}\,{dt}^2
 -\frac{1}{c^2}
 \left(
 \frac{dr^2}{e^{-\frac{r_s}{r}}}
+d\Omega^2
\right)
\right]}=0
\end{equation}

Taking the limit of Eq.(\ref{full}) for $\frac{r_s}{r} <<1$, using the first order approximation for the exponential terms and simplifying we have:

\begin{equation}
\delta\int_{path}
\left[
 c^2\,{dt}^2{(1-\frac{r_s}{r}}) -\frac{{dr}^2}{(1-\frac{r_s}{r})}-d\Omega^2
 \right]
 =0 
 \end{equation} 

 This is equivalent to minimizing the variation of  Eq. (\ref{Schwarzschild_metric}), the Schwarzschild metric which leads to the equations of motion for geodesics in general relativity.   Hence the non-metric theory presented here reduces to general relativistic gravitation for the weak field limit.    The exact form is the Lagrangian of Eq. (\ref{Lagrangian}). 
 
 \section{5. CONCLUSIONS}
Transformations are developed from a non-accelerating reference frame and a frame in free fall in a gravitational field which is also moving with a relative velocity $v$ with respect to the reference frame.   These transformations are everywhere finite and free of singularity except at the radial coordinate origin.  These lead to a Lagrangian that reduces to the Schwarzschild metric in the weak field limit, but that is nonsingular at the Schwarzschild radius.  The theory shows that particles may indeed escape from any initial radial location (to infinity). A particle's energy is reduced by gravitational red shifting  by a factor determined by its initial radial location.  From Eq. (\ref{red_shift}) this factor is $e^{-1}$ at the gravitational radius, and $e^{-\frac{1}{2}}$ at the Schwarzschild radius.  Therefore  black holes are not actually black.
 
The instantaneous effects of gravitation in the theory presented here and general relativity are related by the factors $e^{-\frac{r_g}{r}}$ and $1-\frac{r_g}{r}$ respectively.  A numerical comparison of these factors shows that the two theories become distinguishable in the approximate regime of $r<10 \ r_g$. The current state of the art as indicated by recent direct observations of object motion near the Milky Way's central black hole \cite{Gillessen} and observations of multiple black hole objects with the Event Horizon Telescope  \cite{EHT Pubs} indicate that the required regime is being approached although it has not yet been achieved. Additional methods including cumulative effects (e.g. orbital rotation), or gravitational red shift, or Doppler effects may also be required.

\subsection{}
\subsubsection{}

\section{}
\subsection{}

\subsubsection{}

\section{}
\subsection{}
\subsubsection{}

\section{}
\subsection{}
\subsubsection{}

\begin{acknowledgments}
The author is greatly indebted to Dr. John Anderson without whose critique important corrections and improvements would not have been made.   I am also thankful for numerous helpful and formative discussions with Dr. John S. Scott.

\end{acknowledgments}


\end{document}